\documentclass{Interspeech}
\usepackage[table]{xcolor}
\usepackage{tikz}
\usepackage{pgfplots}
\pgfplotsset{compat=1.10}
\usepackage{todonotes}
\usepackage{float}
\usepackage{booktabs}
\usepackage{pbalance}
\usepackage{upgreek}

\interspeechcameraready

\title{Benchmarking Neural Speech Codec Intelligibility with SITool}

\author[affiliation={1}]{Anna}{Leschanowsky}
\author[affiliation={1}]{Kishor}{Kayyar Lakshminarayana}
\author[affiliation={1}]{Anjana}{Rajasekhar}
\author[affiliation={1}]{Lyonel}{Behringer}
\author[affiliation={2}]{Ibrahim}{Kilinc}
\author[affiliation={1}]{Guillaume}{Fuchs}
\author[affiliation={3}]{Emanuël A. P.}{Habets}

\affiliation{}{Fraunhofer Institute for Integrated Circuits (IIS), Erlangen}{Germany}
\affiliation{Department of Electrical and Computer Engineering}{University of California, San Diego}{USA}
\affiliation{}{International Audio Laboratories Erlangen}{Germany}
\email{anna.leschanowsky@iis.fraunhofer.de}
\keywords{speech coding, speech intelligibility, subjective evaluation, Rhyme Test}

\newcommand{\changed}[1]{{\color{black}{#1}}}%

\usepackage{comment}

\begin{document}

\maketitle

\begin{abstract}

Speech intelligibility assessment is essential for evaluating neural speech codecs, yet most evaluation efforts focus on overall quality rather than intelligibility. Only a few publicly available tools exist for conducting standardized intelligibility tests, like the Diagnostic Rhyme Test (DRT) and Modified Rhyme Test (MRT). We introduce the Speech Intelligibility Toolkit for Subjective Evaluation (SITool), a Flask-based web application for conducting DRT and MRT in laboratory and crowdsourcing settings. We use SITool to benchmark 13 neural and traditional speech codecs, analyzing phoneme-level degradations and comparing subjective DRT results with objective intelligibility metrics. Our findings show that, while neural speech codecs can outperform traditional ones in subjective intelligibility, only STOI and ESTOI -- not WER -- significantly correlate with subjective results, although they struggle to capture gender and wordlist-specific variations observed in subjective evaluations. 
\end{abstract}

\section{Introduction}

Neural speech codecs have gained increasing interest for their ability to outperform traditional coding schemes, especially at very low bitrates~\cite{kim2025neural}. Unlike traditional approaches, they use data-driven methods that avoid costly tuning via listening tests~\cite{kim2025neural} but may hallucinate content due to their generative nature. Thus, performance evaluation remains essential. Evaluation can be subjective (e.g., P.800~\cite{ITUTP800}, MUSHRA~\cite{ITU-BS1534-3-2015}) or objective (e.g., VISQOL~\cite{chinen2020visqol}, ESTOI~\cite{jensen2016algorithm, brendel2024neural, kumar23_dac}). A recent subjective study utilized Degradation Category Ratings (DCR) to benchmark neural codecs~\cite{muller2024speech}. While P.800 and its crowdsourcing variant P.808 are widely used, they assess overall speech quality rather than speech intelligibility. Yet, as neural speech codecs are increasingly deployed in digital communication, broadcasting, and media streaming, intelligibility becomes a crucial evaluation criterion. Speech intelligibility, a distinct dimension of speech quality, is not always strongly correlated with overall quality~\cite{preminger1995quantifying}. Consequently, dedicated intelligibility assessments offer deeper insights into codec performance. 

Speech intelligibility can be assessed at phoneme, word, or sentence level. At the sentence level, Beno{\^i}t et al.~\cite{benoit1996_SUS_SC} proposed a Semantically Unstable Sentence (SUS) test, which measures intelligibility through Word Error Rates (WER) but can be unreliable due to participants' spelling errors. 
Word-level tests such as the Phonetically Balanced Word Test (PBWT), Diagnostic Rhyme Test (DRT), and Modified Rhyme Test (MRT) have been standardized by the American National Standards Institute (ANSI)~\cite{ANSI}. PBWT is an open-set test requiring transcription of a word, while DRT and MRT are closed-set tests where listeners select the heard word from a set of two (DRT) or six (MRT) options. Closed-set tests are easier to administer, require fewer listeners, and show high reliability, whereas open-set tests are less prone to ceiling effects~\cite{schmidt1995intelligibility}. The DRT consists of 96 rhyming word pairs that differ only in their initial or final consonant sounds~\cite{ITUTP807}. A key advantage of DRT is that it allows for detailed analysis of phoneme-specific degradations by providing diagnostic feature scores across six categories~\cite{ITUTP807}.

Crowdsourced listening tests are widely adopted for speech quality~\cite{ITUTP808} and, more recently, for intelligibility evaluation~\cite{lechler2024_icassp_crowd_si}. While open-source tools for speech quality assessment exist (e.g., WebMUSHRA~\cite{SchoefflerEtAl15_webMUSHRA_WAC}, HultiGen-v2\footnote{\url{https://github.com/APL-Huddersfield/HULTI-GENv2/}}), open-source DRT and MRT software toolkits remain limited or outdated~\cite{calistri1986drt}. For instance, the scripts for deploying DRT, provided in \cite{lechler2024_icassp_crowd_si}, are restricted to Qualtrics and do not support MRT, whereas the Python-based MRT tool provided by NIST is not optimized for crowdsourced test environments\footnote{\url{https://github.com/usnistgov/MRT}}. 

To fill this gap, we introduce SITool -- a Flask-based web application for DRT and MRT -- and benchmark 13 speech codecs with English male and female talkers. An overview of benchmarked codecs is given in Section~\ref{sec:SpeechCodecs}, and the tool is described in Section~\ref{sec:SITool}. 
While subjective tests remain the gold standard, the correlation between objective intelligibility metrics (e.g., STOI, ESTOI, and WER~\cite{jensen2016algorithm, TaalSTOI, KARBASI2022108606}) and subjective DRT scores for neural speech codecs is underexplored. Therefore, we present subjective and objective results and analyze their relationship in Section~\ref{sec:Results}.




\section{Benchmarked Speech Codecs}
\label{sec:SpeechCodecs}

We evaluate the intelligibility of various publicly available state-of-the-art neural codecs and, for a full performance evaluation, compare them with prominent traditional speech codecs.

In this work, we consider \textbf{LPCNet}\footnote{\url{https://github.com/xiph/LPCNet/commit/7dc9942}}~\cite{valin2019lpcnet} at 1.6 kbps, an early solution that combines classical signal processing and deep neural networks to decode a bitstream generated by a conventional encoder. The hybrid solution allows an autoregressive neural vocoder to run in real-time on a CPU (3 GFLOPs). Further, we evaluate different GAN-based end-to-end autoencoder approaches, a paradigm introduced by SoundStream~\cite{zeghidour2022_soundstream} built on the discrete learned representation of VQ-VAE~\cite{van_den_oord_2017}. This principle is adopted by most state-of-the-art neural speech codecs. We test \textbf{Lyra V2}\footnote{\url{https://github.com/google/lyra/tree/v1.3.2}}~\cite{lyra}, an open-source codec derived directly from the original SoundStream model and optimized to run in real-time on a smartphone CPU. Various improvements to the paradigm have been proposed by adding recurrent layers~\cite{defossez2023_encodec}, increasing the model size~\cite{kumar23_dac}, or carefully selecting the training data~\cite{shechtman24_dac_ibm}.  It has also been proposed to modify them by exploiting the strong generative capacity of diffusion models~\cite{yang24_ladiffcodec,liu24_semanticodec}, or by semantically disentangling the latent space to achieve even lower bitrates~\cite{liu24_semanticodec, defossez2024_moshi}. Transformers and attention layers have also been considered to get an even more compact representation~\cite{defossez2024_moshi, siahkoohi22_interspeech}. Among them, we evaluate two extensions of DAC~\cite{kumar23_dac}, namely a speech fine-tuned version, \textbf{DAC-S}~\cite{shechtman24_dac_ibm}, at 1.5 kbps, and the multi-scale quantization \textbf{SNAC}~\cite{siuzdak2024snac}, at 980 bps. 
\textbf{Mimi}~\cite{defossez2024_moshi}, evaluated at 0.55 and 1.1 kbps, is characterized by transformer bottlenecks before and after quantization, as well as the inclusion of semantic tokens via distillation. We also include the diffusion-based \textbf{SemantiCodec}~\cite{liu24_semanticodec}, which combines acoustic and semantic encoders. With 507M parameters, it is the largest of the codecs evaluated, allowing for very low bitrates (340 and 680 bps).

For comparison purposes, three generations of traditional 3GPP communication codecs, \textbf{AMR}~\cite{amr}, \textbf{AMR-WB}~\cite{amr-wb}, and \textbf{EVS}~\cite{evs} are included in the study at 4.75, 6.6 and 8 kbps, respectively. They are all based on the popular CELP paradigm, a hybrid approach aiming at preserving the waveform through an analysis-by-synthesis optimization of its coded parameters. We employ them at, or near, their minimum bitrate. In addition, the purely parametric \textbf{Codec2}~\cite{codec2} speech coder is considered at its minimum bitrate of 0.7 and at 2.4 kbps.

\section{SITool - Speech Intelligibility Toolkit for Subjective Evaluation}
\label{sec:SITool}

We developed a standalone Flask-based web application that supports both DRT and MRT testing and can be deployed in laboratory settings or on crowdsourcing platforms like Amazon Mechanical Turk (AMT)\footnote{We make the framework publicly available at \url{https://github.com/audiolabs/SITool}}. Our tool allows the integration of consent forms, language proficiency assessments, predefined trap questions, and enforced breaks through an easily customizable configuration file. The application outputs detailed results files, which automatically analyze for accuracy across test conditions and, in the case of DRT, across acoustic features. Additionally, the toolkit includes scripts for further analysis.

\subsection{Word Lists, Speech Samples and Test Blocks}

For benchmarking, we used English word lists from~\cite{ITUTP807} and audio stimuli provided by~\cite{lechler2024_icassp_crowd_si}. Testing was done at 16~kHz sampling rate. Audio for codecs operating at a different sampling rate was resampled before coding and converted back to 16 kHz afterward\footnote{Some examples are provided at \url{https://s.fhg.de/sitoolweb}}. To reduce test time, we randomly selected male and female talker samples from the original dataset. Importantly, talkers varied only within and across word lists but remained constant across codecs. Word pairs were divided into four sessions~\cite{ITUTP807}, resulting in a total of eight sessions.

\subsection{Trap Questions, Gold Standard and Lower Anchor}
\label{sec:trapQuestions}

Following best practices for crowdsourced listening tests~\cite{ITUTP808}, the DRT included trap questions, the original signal as a gold standard, and a lower anchor for post-screening purposes. Trap questions consisted of two phonetically distinct words from the DRT word list to ensure participants' attentiveness. The clean speech signal served as a reference for codec performance and ensured that participants could reliably identify clean recordings. The lower anchor was designed as a control condition for unintelligibility and reliability of the interface to ensure no unintended cues helped with word recognition. Therefore, we synthesized words without their distinguishing consonants using a synthesis system consisting of an acoustic model based on ForwardTacotron\footnote{\url{https://github.com/as-ideas/ForwardTacotron}}, with extensions similar to~\cite{ZalkowEtAl23_AudioLabs_Blizzard} and StyleMelGAN as neural vocoder~\cite{MustafaPF21_StyleMelGAN_ICASSP}. 
To entirely obscure the incomplete word and ensure no residual information could help to distinguish between words, we added cough noise from~\cite{Patel_Rivas_Psaltos_2020} to the location where the consonant was removed.

\subsection{Test Procedure}

\begin{figure}[!t]
    \centering
    \includegraphics[width=0.6\columnwidth]{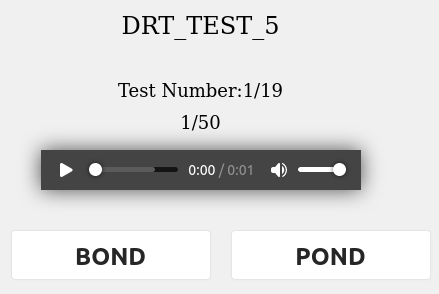}
    \caption{Graphical user interface for the DRT test}
    \vspace{-0.3cm}
    \label{fig:gui}
\end{figure}

Prior to the test, informed consent was obtained, and demographic information on age, gender, and self-assessed language proficiency was collected. In addition, to ensure a sufficient level of English language proficiency, participants completed a B1/B2 Level test based on official Cambridge English Listening Test examples\footnote{\url{https://www.cambridgeenglish.org/learning-english/activities-for-learners/}}. To qualify for the main test, they had to answer at least four out of five questions correctly and were instructed to use headphones. After passing, participants completed a trial run to familiarize themselves with the test interface (see Figure~\ref{fig:gui}). The main test included 15 conditions from either male or female talkers and was split into two parts of roughly 40 minutes with a mandatory five-minute break in between.

\subsection{Participants and Post-Screening}

Participants were recruited from AMT and an internal listener pool with an average hourly compensation of \$10.50 USD, aligning with prior crowdsourcing studies in this context~\cite{lechler2024_icassp_crowd_si}. For the final sample, we excluded participants who failed any trap questions but not based on their lower anchor performance, as no outliers were identified. The final sample consisted of 82 participants, unevenly distributed across the eight tests, but with at least eight participants per test as recommended by~\cite{ITUTP807}. Of these, 54.88\% identified as male, 43.90\% as female, and 1.22\% as other. The sample was diverse in age (M=43.74, SD=10.68), ranging from 23 to 70 years. 66 participants were native English speakers (US: 56, UK: 3, Other: 7), 16 non-native speakers. 



\begin{figure}[t]
    \centering
    \includegraphics[width=\columnwidth]{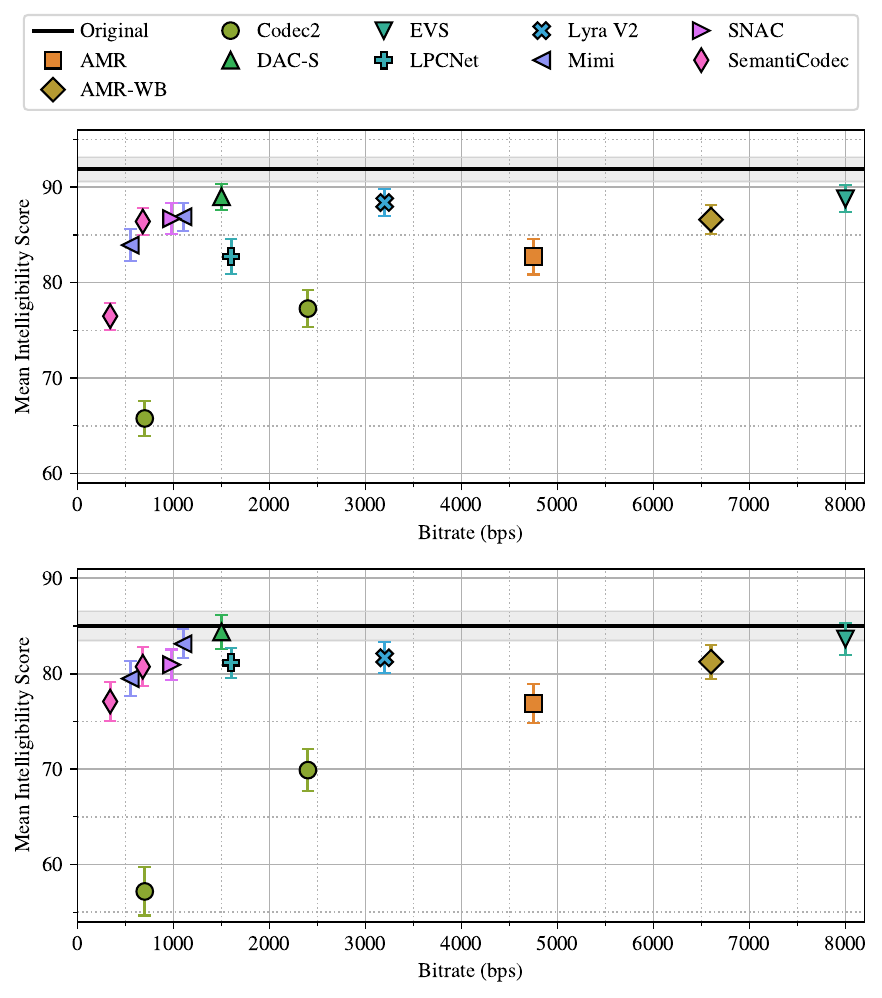}
    \vspace{-1em}
    \caption{Mean subjective intelligibility scores per talker gender, with confidence intervals by bitrate (top: male, bottom: female). The low anchor (mean of 2.98 for male and -1.46 for female) is omitted for clarity.}
    \label{fig:subj_score_by_bps}
\end{figure}

\subsection{Score Calculation}

Intelligibility scores were calculated according to~\cite{ITUTP807} and adjusted for guessing with $R$ being the number of correct answers, $W$ being the number of incorrect answers: $P(c) = \frac{R-W}{R+W} * 100$.

\section{Results}
\label{sec:Results}

\begin{table*}[!h]
    \centering
    \caption{List of tested codecs, along with intelligibility objective metrics. 45 items were too short for STOI/ESTOI computation and were excluded. Confidence intervals are omitted since they are almost negligible, all under 0.015.}
    \footnotesize
    \resizebox{0.65\textwidth}{!}{%
    \begin{tabular}{l c c c c c c}
        \toprule
        \textbf{Codec} & \textbf{Params (M)} & \textbf{kHz} & \textbf{kbps} & \textbf{STOI}  f/m ($\uparrow$) & \textbf{ESTOI} f/m 
 ($\uparrow$) &  \textbf{WER} f/m ($\downarrow$)\\
        \midrule
        Original & - & - & - & 1.00 / 1.00 & 1.00 / 1.00  & 0.25 / 0.19 \\
        \midrule
        Codec2~\cite{codec2} & - & 8 & 0.7 & 0.57 / 0.60 &  0.44 / 0.39 & 0.86 / 0.81\\
        Codec2~\cite{codec2} & - & 8 & 2.4 & 0.75 / 0.74  &  0.63 / 0.52 & 0.54 / 0.54\\
        AMR~\cite{amr} & - & 8 & 4.75 & 0.87 / 0.88 & 0.80 / 0.75 & 0.43 / 0.39\\
        AMR-WB~\cite{amr-wb} & - & 16 & 6.6 & 0.90 / 0.90 & 0.83 / 0.78 & 0.30 / 0.28 \\
        EVS~\cite{evs}  & -  & 16 & 8 &  \textbf{0.95 / 0.94} & \textbf{0.91 / 0.87} & \textbf{0.29 / 0.30}\\
        \midrule
        LPCNet~\cite{valin2019lpcnet} & 0.073 & 16 & 1.6 & 0.78 / 0.80 & 0.67 / 0.60 &  0.46 / 0.55 \\
        Lyra V2~\cite{lyra} & n/a & 16 & 3.2 & 0.90 / 0.89 & 0.82 / 0.75 &  0.38 / 0.35\\
        DAC-S~\cite{shechtman24_dac_ibm} & 76 & 24 & 1.5 & \textbf{0.93 / 0.92} & \textbf{0.86 / 0.79} & \textbf{0.34 / 0.31}\\
        SNAC~\cite{siuzdak2024snac} & 20 & 24 & 0.98 & 0.89 / 0.89 & 0.80 / 0.72 & 0.44 / 0.47 \\
        Mimi~\cite{defossez2024_moshi} & 80 & 24 & 0.55 & 0.85 / 0.86 & 0.73 / 0.65 & 0.51 / 0.47\\
        Mimi~\cite{defossez2024_moshi} & 80 & 24 & 1.1 & 0.91 / 0.91 & 0.83 / 0.76 & 0.35 / 0.35\\
        SemantiCodec~\cite{liu24_semanticodec} & 507 & 16 & 0.34 & 0.79 / 0.80 & 0.64 / 0.55&  0.68 / 0.69\\ 
        SemantiCodec~\cite{liu24_semanticodec} & 507& 16 & 0.68 & 0.84 / 0.85 & 0.72 / 0.65 &  0.49 / 0.44\\ 
        \bottomrule
    \end{tabular}
    }
    \label{tab:obj_results}
\end{table*}

\subsection{Subjective Speech Intelligibility Evaluation}

\begin{figure*}[!ht]
    \centering
    \includegraphics[width=0.8\textwidth]{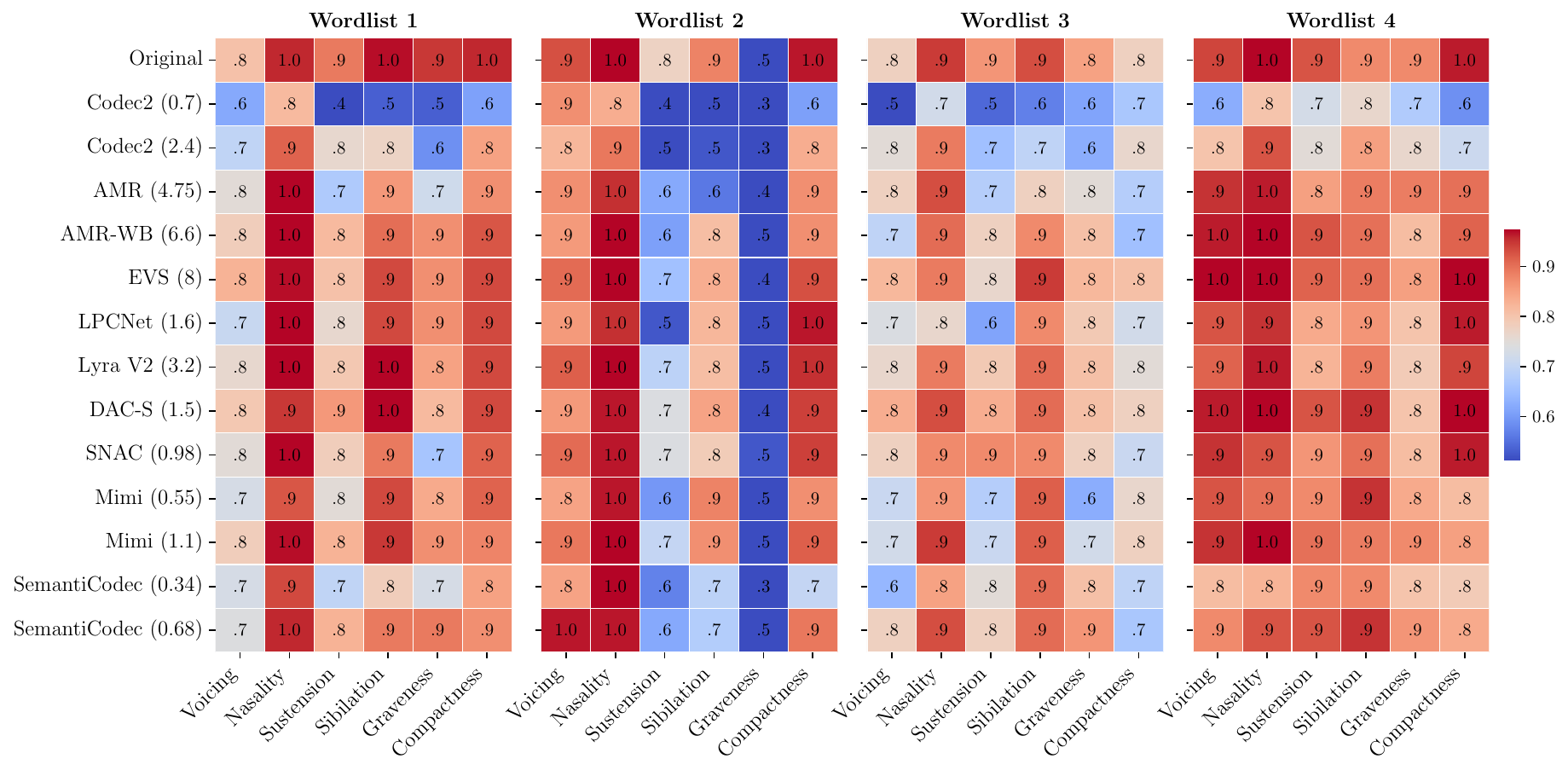}
    \vspace{-1em}
    \caption{Accuracy heatmap for codecs across distinctive features and wordlists as outlined in~\cite{ITUTP807}. The x-axis shows the distinctive features and the y-axis shows the codecs with their respective bitrates (kbps) in brackets.}
    \label{fig:heatmap}
\end{figure*}


Subjective intelligibility results for male and female talkers are shown in Figure~\ref{fig:subj_score_by_bps}. The results show that the female reference signal (M=85, SD=9.66) has a noticeably lower intelligibility score than the male reference signal (M=91.87, SD=8.23). This pattern can be observed across most codecs, except for SemantiCodec at 340 bps, where female talkers slightly outperformed males (77.8\% vs. 76.49\%). Moreover, despite operating at a lower bitrate, DAC-S slightly outperformed EVS for both genders. The lower anchor results (M=0.81 SD=10.52) confirmed its effectiveness as an unintelligibility benchmark and are excluded from further analysis.

To assess performance differences across codec conditions and talker gender, we employed a Linear Mixed-Effects Model (LMM) with fixed effects for codec condition (14 levels), talker gender (2 levels), and word list (4 levels) and random effects for individual participants to account for within-subject variability. A likelihood-ratio test showed that adding interaction effects for all factors significantly improved the model fit compared to a purely additive model ($\chi^2$(94)=217.48, p$<$.0001). Visual inspection of residual plots showed no clear deviations from homoscedasticity or normality. The model's fixed effect intercept was 87.5 [82.32, 92.68]. A repeated measures ANOVA (type III, Satterthwaite approximation of degrees-of-freedom) on a restricted maximum likelihood estimation (REML) refitted model revealed significant main effects for codec condition (F(13, 1065.68)=98.10, p$<$.0001), talker gender (F(1, 79.71)=12.59, p$<$.0001), and wordlist (F(3, 89.65)=17.31, p$<$.0001) as well as significant interaction effects of condition-gender (F(13, 1065.68)=3.035, p$<$.0001), condition-wordlist (F(39, 1065.68)=2.96, p$<$.0001) and gender-wordlist (F(3, 89.65)=16.30, p$<$.0001). \changed{Yet, wordlist differences might not be reliably detectable with the current sample and should be confirmed with a larger one.} Post-hoc tests using ``emmeans''~\footnote{\url{https://rvlenth.github.io/emmeans/}} showed that, when averaged over gender and word lists, all codecs except EVS and DAC-S performed significantly worse than the reference (p$<$.01). However, interaction effects revealed that most codecs differed significantly from the reference for male talkers, whereas only Codec~2, AMR, Mimi at 550~bps and SemantiCodec at 340~bps scored significantly lower for females, likely due to consistently lower intelligibility scores for the female references. Although the crowdsourced audio samples were carefully curated~\cite{lechler2024_icassp_crowd_si}, these gender-specific differences for the reference signal warrant further investigation. Future studies should explicitly control for talker-specific effects and assess the potential impact of consumer recording devices on crowdsourced recordings for intelligibility assessment~\cite{busquet2024voice}.

Further, we found that intelligibility scores for male talkers outperformed those of female talkers on Wordlist~1,~2 and~4 for all codecs but not on Wordlist~3. Wordlist~4 yielded highest intelligibility scores for both genders and all codecs, whereas Wordlist~2 performed worst for females and second worst for male talkers. \changed{These gender-specific differences already exist at the reference level and are not due to codec performances. Even when using curated datasets as in this test, it is essential to include reference signals and possibly conduct pre-tests on the intelligibility of reference signals.} A heatmap (Figure~\ref{fig:heatmap}) shows that the distinctive feature ``graveness'' in Wordlist~2 had notably low scores across codecs, including the reference. Closer inspection revealed that low performance was linked to high error rates for word pairs `TROUGH-TROTH' and `FIN-THIN', which repeatedly test the consonants /f/ and /$\uptheta$/. Similarly, Wordlist~4 showed more errors for the word pair `REEF-WREATH', with little impact on average performance. We did not observe similar patterns for word pairs of ``graveness'' including consonants /f/ and /$\uptheta$/ in Wordlist~3, suggesting that both phoneme characteristics and individual pronunciation influenced the results.

\subsection{Objective Speech Intelligibility Evaluation}

We show ojective results\footnote{WER calculated using Whisper ``large'' from \url{https://github.com/openai/whisper/}.} in Table~\ref{tab:obj_results}. First, we observe that WER is rather high even for the reference signals, which might indicate that single-word utterances pose a challenge to Whisper. 
Across all codecs, female ESTOI scores are at least 0.04 higher than male ones, while no such trend is visible for STOI and WER. Looking at traditional and neural codecs separately, EVS and DAC-S show the best results. However, while DAC-S outperforms EVS and AMR-WB in the subjective tests, it surpasses EVS in all objective metrics and AMR-WB in WER. Further, SemantiCodec at 680 bps ranks below Mimi at 550 bps w.r.t. STOI and ESTOI, differing from the subjective results.

Performance differences were evaluated across codec conditions, gender, and word lists using two linear regression models. For STOI and ESTOI, additive models were chosen over interaction models because they explained equivalent variance ($R^2_{STOI}=0.79$,$R^2_{ESTOI}=0.71$) while showing lower AIC and BIC values. 
A Type III ANOVA on the STOI model revealed significant effects of codec condition (F(13) = 1349.2, p$<$.0001) and word list (F(3) = 28.82, p$<$.0001), but not gender. In contrast to subjective results, pairwise comparisons indicated that all codecs were significantly different from the reference, with Wordlist~4 showing the lowest and Wordlist~2 and~3 the highest performances. For ESTOI, condition (F(13)=861.15, p$<$.0001), gender (F(1)=564.89, p$<$.0001), and word list (F(3)=11.49, p$<$.0001) all significantly affected scores. However, unlike subjective results, female talkers obtained higher scores across all word lists and codec conditions than male talkers. Finally, fitting a model to WER yielded results similar to those for STOI, indicating that objective measures are misaligned with subjective performance when considering subtle differences between gender and word lists.

\subsection{Correlation Analysis}

We conducted a correlation analysis on the mean subjective and objective results disaggregated by codecs, wordlists, and talker gender. The analysis showed that the correlation between subjective and objective metrics -- STOI and ESTOI -- was considerably higher when averaging over gender and wordlist. For STOI, Pearson's r increased from 0.595 to 0.958, and for ESTOI from 0.499 to 0.890. This, in combination with the previous analysis, suggests that the objective metrics do not fully capture the variability of talker gender and wordlist obtained in the DRT. In contrast, subjective results and WER were uncorrelated (non-averaged: -0.11, averaged: -0.15), questioning the usefulness of WER as an objective metric for speech intelligibility in the context of DRT.


\vspace{-1em}
\section{Conclusions}

We benchmarked 13 speech codecs, from traditional to state-of-the-art GAN-based codecs, on intelligibility using the newly developed \textit{SITool}. Unlike speech quality evaluation, DRT offered a more detailed analysis of phoneme-specific degradations and distinctive acoustic features. 
Our results show that below 1100 bps, the intelligibility of neural codecs degrades compared to the uncoded signal, leaving room for potential improvement. Notably, the female reference signal yielded overall lower intelligibility scores, which may have impacted the evaluation of codecs and calls for further investigation of audio material. While objective measures, i.e., STOI and ESTOI, correlate well with subjective intelligibility when averaged over talker gender and word lists, they fail to capture finer variations related to gender and wordlist differences identified in the subjective evaluation, and WER showed no correlation with subjective results. As neural codecs improve, subjective tests remain crucial for detecting subtle impairments. Given the observed ceiling effects, future work should explore open-set tests like SUS, less prone to ceiling effects \changed{and not reliant on wordlists}~\cite{schmidt1995intelligibility}.


\bibliographystyle{IEEEtran}

\bibliography{mybib}

\end{document}